\documentclass{PoS}

\title{Definitions of a static SU(2) color triplet potential}

\ShortTitle{Definitions of a static SU(2) color triplet potential}

\author{\speaker{Marc Wagner} \\
        Goethe-Universit\"at Frankfurt am Main, Institut f\"ur Theoretische Physik, \\ Max-von-Laue-Stra{\ss}e 1, D-60438 Frankfurt am Main, Germany \\
        E-mail: \email{mwagner@th.physik.uni-frankfurt.de}}

\author{Owe Philipsen \\
        Goethe-Universit\"at Frankfurt am Main, Institut f\"ur Theoretische Physik, \\ Max-von-Laue-Stra{\ss}e 1, D-60438 Frankfurt am Main, Germany \\
        E-mail: \email{philipsen@th.physik.uni-frankfurt.de}}

\abstract{We discuss possibilities and problems to non-perturbatively define and compute a static color triplet potential in SU(2) gauge theory. Numerical lattice results are presented and compared to analytical perturbative results.
}

\FullConference{Xth Quark Confinement and the Hadron Spectrum\\
                8-12 October 2012 \\
                TUM Campus Garching, Munich, Germany}

\begin{document}



\hspace{-0.7cm}\textbf{Calculating the static potential in SU(2) gauge theory: basic principle}

The calculation of the singlet static potential is usually based on trial states \\ $| \Phi^\textrm{\scriptsize sing} \rangle \equiv \bar{Q}(-r/2) U(-r/2;+r/2) Q(+r/2) | \Omega \rangle$, while for the triplet static potential typically \\ $| \Phi^{\textrm{\scriptsize trip},a} \rangle \equiv \bar{Q}(-r/2) U(-r/2;s) \sigma^a U(s;+r/2) Q(+r/2) | \Omega \rangle$ is suggested or used (cf.\ e.g.\ \cite{Brambilla:1999xf}). Here $\pm r/2 \equiv (0,0,\pm r/2)$, $Q$ and $\bar{Q}$ are static quark/antiquark operators, $U$ are spatial parallel transporters (on a lattice products of links) and $\sigma^a$ denote Pauli matrices acting in color space. From the asymptotic behavior of the corresponding temporal correlation function the static potential $V_0^X(r)$, $X \in \{ \textrm{singlet}, \textrm{triplet}\}$ can be extracted.
\vspace{0.3cm}



\hspace{-0.7cm}\textbf{Lattice computations without gauge fixing}

On the lattice the singlet correlation function is proportional to Wilson loops, \\ $\langle \Phi^\textrm{\scriptsize sing}(t_2) | \Phi^\textrm{\scriptsize sing}(t_1) \rangle \propto W(r,\Delta t)$, $\Delta t = t_2 - t_1$, from which the singlet potential can be determined (cf.\ the figure on page 1, blue dots). Since the triplet correlation function is not gauge invariant, one obtains $\langle \Phi^{\textrm{\scriptsize trip},a}(t_2) | \Phi^{\textrm{\scriptsize trip},a}(t_1) \rangle = 0$ and cannot determine a triplet potential.
\vspace{0.3cm}



\hspace{-0.7cm}\textbf{Lattice computations in temporal gauge}

Temporal gauge $A_0^g = 0$ in the continuum corresponds to temporal links $U_0^g(t,\mathbf{x}) = 1$ on a lattice. These links gauge transform according to $U_0(t,\mathbf{x}) \rightarrow U_0^g(t,\mathbf{x}) = g(t,\mathbf{x}) U_0(t,\mathbf{x}) g^\dagger(t+a,\mathbf{x})$, where $g(t,\mathbf{x}) \in \textrm{SU(2)}$. On a lattice with finite periodic temporal extension it is not possible to realize temporal gauge everywhere. There will be a slice of links, where $U_0 \neq 0$ (in the following wlog.\ $U_0^g(t=0,\mathbf{x}) \neq 1$, while $U_0^g(t=1 \ldots T-1,\mathbf{x}) = 1$; $T$ is the periodic temporal extension of the lattice). A possible choice for the corresponding gauge transformation $g(t,\mathbf{\mathbf{x}})$ is \\
$g(t=2a,\mathbf{x}) = U_0(t=a,\mathbf{x})$, \\
$g(t=3a,\mathbf{x}) = g(t=2a,\mathbf{x}) U_0(t=2a,\mathbf{x}) = U_0(t=a,\mathbf{x}) U_0(t=2a,\mathbf{x})$, \\
$g(t=4a,\mathbf{x}) = g(t=3a,\mathbf{x}) U_0(t=3a,\mathbf{x}) = U_0(t=a,\mathbf{x}) U_0(t=2a,\mathbf{x}) U_0(t=3a,\mathbf{x})$, $\ldots$
\vspace{0.3cm}

\hspace{-0.7cm}\textit{Non-perturbative computations (lattice), singlet potential:}

The trial states $| \Phi^\textrm{\scriptsize sing} \rangle$ are gauge invariant. Therefore, the result is identical to the result without gauge fixing (cf.\ the figure on page~1, blue dots).

Gauge transforming the temporal links to $U_0^g(t,\mathbf{x}) = 1$ and computing
\begin{eqnarray}
\langle \Phi^\textrm{\scriptsize sing}(t_2) | \Phi^\textrm{\scriptsize sing}(t_1) \rangle \ \ = \ \ \Big\langle \textrm{Tr}\Big(U^g(t_1,-r/2;t_1,+r/2) U^g(t_2,+r/2;t_2,-r/2)\Big) \Big\rangle
\end{eqnarray}
(here we assume $1 \leq t_1 < t_2 < T$, ``case (A)'') is equivalent to consider the manifestly gauge invariant observable
\begin{eqnarray}
\nonumber & & \hspace{-0.7cm} \langle \Phi^\textrm{\scriptsize sing}(t_2) | \Phi^\textrm{\scriptsize sing}(t_1) \rangle \ \ = \ \ \Big\langle \textrm{Tr}\Big(U(t_1,-r/2;t_1,+r/2) \underbrace{g^\dagger(t_1,+r/2) g(t_2,+r/2)}_{U(t_1,+r/2;t_2,+r/2)} \\
 & & \hspace{0.675cm} U(t_2,+r/2;t_2,-r/2) \underbrace{g^\dagger(t_2,-r/2) g(t_1,-r/2)}_{U(t_2,-r/2;t_1,-r/2)}\Big) \Big\rangle \ \ = \ \ W(r,\Delta t)
\end{eqnarray}
(cf.\ the figure on page~1). Similar considerations yield the same result for ``case (B)'', $0 = t_1 < t_2 < T$ or $1 \leq t_2 < t_1 < T$. This technique of transforming a non-gauge invariant observable into an equivalent manifestly gauge invariant observable will be helpful for interpreting the triplet potential.

A helpful theoretical tool to understand, which states are contained in a correlation function, is the transfer matrix formalism (cf.\ e.g.\ \cite{Philipsen:2001ip,Jahn:2004qr}). Without gauge fixing the transfer matrix is $\hat{T} = e^{-H a}$, $\hat{T} | \psi^{(n)} \rangle = \lambda^{(n)} | \psi^{(n)} \rangle$, $\lambda^{(n)} = e^{-E^{(n)} a}$ (lattice discretization errors neglected), where $E^{(n)}$ are the energies of gauge invariant states (e.g.\ the vacuum, glueballs). Similarly the transfer matrix in temporal gauge is $\hat{T}_0 = e^{-H_0 a}$, $\hat{T}_0 | \psi_0^{(n)} \rangle = \lambda_0^{(n)} | \psi_0^{(n)} \rangle$. In temporal gauge remaining gauge degrees of freedom are time-independent gauge transformations $g(\mathbf{x})$. One can show $[\hat{T}_0,g(\mathbf{x})] = 0$, i.e.\ eigenstates of $\hat{T}_0$ can be classified according to SU(2) color quantum numbers $(j(\mathbf{x}),m(\mathbf{x}))$ at each $\mathbf{x}$. $\lambda_0^{(n)} = e^{-E_0^{(n)} a}$, where $E_0^{(n)}$ are the energies of the gauge invariant states already mentioned as well as of additional non-gauge invariant states with $j(\mathbf{x}) \neq 0$. Such states can be interpreted as states containing static color charges (= static quarks)\footnote{We use the following notation of energy eigenvalues $E_0^{(n)}$: (1) gauge invariant states, i.e.\ no static quarks: $\mathcal{E}_n$ ($j(\mathbf{x}) = 0$ for all $\mathbf{x}$); (2) a static quark/antiquark at $-r/2$ and at $+r/2$: $V^\textrm{\scriptsize sing}_n(r)$ ($j(-r/2) = j(+r/2) = 1/2$); (3) an adjoint static quark at $s$: $\mathcal{E}_n^{Q^\textrm{\scriptsize adj}}$ ($j(s) = 1$); (4) a static quark/antiquark at $-r/2$ and at $+r/2$, an adjoint static quark at $s$: $V^{Q \bar{Q} Q^\textrm{\scriptsize adj}}_n(r)$ ($j(-r/2) = j(+r/2) = 1/2$, $j(s) = 1$).}. One can derive
\begin{eqnarray}
\langle \Phi^\textrm{\scriptsize sing}(t_2) | \Phi^\textrm{\scriptsize sing}(t_1) \rangle \ \ = \ \ \sum_k e^{-V_k^\textrm{\scriptsize sing}(r) \Delta t} \sum_m e^{-\mathcal{E}_m (T - \Delta t)} \sum_{\alpha,\beta} \Big|\langle k , \alpha \beta | \hat{U}_{\alpha \beta}(-r/2;+r/2) | m \rangle\Big|^2 ,
\end{eqnarray}
where $\alpha \equiv m(-r/2) = \pm 1/2$ and $\beta \equiv m(+r/2) = \pm 1/2$ are color indices at $\pm r/2$. As expected this correlation function is suited to extract the common singlet potential $V_0^\textrm{\scriptsize sing}(r)$.
\vspace{0.3cm}

\hspace{-0.7cm}\textit{Non-perturbative computations (lattice), triplet potential:}

Again one has to distinguish the two cases (A) and (B), which this time yield different results. When including the gauge fixing in the observable, one finds that $(s,t_1)$ and $(s,t_2)$, the spacetime positions of the ``triplet generators'' $\sigma^a$, are connected by an adjoint static propagator: $\textrm{Tr}(\sigma^a U(t_1,s;t_2,s) \sigma^b U(t_2,s;t_1,s))$. Within the transfer matrix formalism one can derive for case (A)
\begin{eqnarray}
\nonumber & & \hspace{-0.7cm} \langle \Phi^{\textrm{\scriptsize trip},a}(t_2) | \Phi^{\textrm{\scriptsize trip},a}(t_1) \rangle \ \ = \\
 & & \hspace{0.675cm} \sum_{\alpha,\beta} \Big|\langle k , \alpha \beta , m(s)=a | \hat{U}_{\alpha \beta,a}(-r/2;s;+r/2) | m \rangle\Big|^2
\end{eqnarray}
and for case (B)
\begin{eqnarray}
\nonumber & & \hspace{-0.7cm} \langle \Phi^{\textrm{\scriptsize trip},a}(t_2) | \Phi^{\textrm{\scriptsize trip},a}(t_1) \rangle \ \ = \ \sum_k e^{-V^\textrm{\scriptsize sing}_k(r) \Delta t} \sum_m e^{-\mathcal{E}_m^{Q^\textrm{\scriptsize adj}} (T - \Delta t)} \\
 & & \hspace{0.675cm} \sum_{\alpha,\beta} \Big|\langle k , \alpha \beta | \hat{U}_{\alpha \beta,a}(-r/2;s;+r/2) | m , m(s)=a \rangle\Big|^2 .
\end{eqnarray}
The conclusion is that one can either extract a three-quark potential (one quark at $+r/2$, one antiquark at $-r/2$, one adjoint quark at $s$) (case (A)) or the ordinary singlet potential (case (B)).
\vspace{0.3cm}



\hspace{-0.7cm}\textbf{Perturbative calculations in Lorenz gauge}

Most perturbative calculations of the static potential are carried out in Lorenz gauge $\partial_\mu A_\mu = 0$. The leading order result for trial states $| \Phi^\textrm{\scriptsize sing} \rangle$ is $V_0^\textrm{\scriptsize sing}(r) = -3 g^2 / 16 \pi r$, i.e.\ an attractive singlet potential. This result can be compared to the non-perturbative lattice result (in any gauge), since the trial state is gauge invariant. To perform a precise matching of lattice and perturbative static potentials, higher orders (NNLO or NNNLO) are required (cf.\ e.g.\ \cite{Jansen:2011vv,Bazavov:2012ka} for recent work on this topic), but nevertheless qualitative agreement is found (cf.\ the figure on page~1, blue dots and blue line). The leading order result for trial states $| \Phi^{\textrm{\scriptsize trip},a} \rangle$ is $V_0^{\textrm{\scriptsize trip}}(r) = +g^2 / 16 \pi r$, i.e.\ a repulsive triplet potential. Note, however, that in Lorenz gauge a transfer matrix does not exist, which renders a physical interpretation difficult. One can also calculate the gauge invariant triplet diagram obtained by using temporal gauge (cf.\ the figure on page~1, ``triplet, case (A)''). Then one obtains \\ $V_0^{Q \bar{Q} Q^\textrm{\scriptsize adj}}(r) = -9 g^2 / 16 \pi r$ (for $s = 0$), i.e.\ an attractive three-quark potential. Again qualitative agreement with the lattice result is found (cf.\ the figure on page~1, red dots and red line).
\vspace{0.3cm}



\hspace{-0.7cm}\textbf{Conclusions}

The singlet potential corresponds to a gauge invariant trial state \\ $\bar{Q}(-r/2) U(-r/2;+r/2) Q(+r/2) | \Omega \rangle$. It is the same in any gauge and its interpretation as a static quark antiquark potential is clear.

The triplet potential corresponding to trial states \\ $\bar{Q}(-r/2) U(-r/2;s) \sigma^a U(s;+r/2) Q(+r/2) | \Omega \rangle$ is different, when using different gauges: (1) without gauge fixing it cannot be calculated/computed; (2) in temporal gauge it corresponds to a three-quark potential and not to a potential between a quark and an antiquark in a color triplet state, i.e.\ the name ``triplet potential'' is misleading; (3) in Lorenz gauge a perturbative calculation yields a repulsive potential; since a transfer matrix does not exist, the physical interpretation is unclear.
\vspace{0.3cm}



\hspace{-0.7cm}\textbf{Acknowledgements}

We thank Felix Karbstein for discussions. M.W.\ acknowledges support by the Emmy Noether Programme of the DFG (German Research Foundation), grant WA 3000/1-1. This work was supported in part by the Helmholtz International Center for FAIR within the framework of the LOEWE program launched by the State of Hesse.
\vspace{0.3cm}




\begin{thebibliography}{99}

\bibitem{Brambilla:1999xf} 
  N.~Brambilla, A.~Pineda, J.~Soto and A.~Vairo,
  Nucl.\ Phys.\ B {\bf 566}, 275 (2000)
  [hep-ph/9907240].

\bibitem{Philipsen:2001ip} 
  O.~Philipsen,
  Nucl.\ Phys.\ B {\bf 628}, 167 (2002)
  [hep-lat/0112047].

\bibitem{Jahn:2004qr} 
  O.~Jahn and O.~Philipsen,
  Phys.\ Rev.\ D {\bf 70}, 074504 (2004)
  [hep-lat/0407042].

\bibitem{Jansen:2011vv} 
  K.~Jansen {\it et al.} [ETM Collaboration],
  JHEP {\bf 1201}, 025 (2012)
  [arXiv:1110.6859 [hep-ph]].

\bibitem{Bazavov:2012ka} 
  A.~Bazavov, N.~Brambilla, X.~Garcia i Tormo, P.~Petreczky, J.~Soto and A.~Vairo,
  arXiv:1205.6155 [hep-ph].

\end{thebibliography}
\end{document}